\documentclass[12pt]{article}
\usepackage{graphicx} % Required for inserting images
\usepackage{subcaption}
\usepackage[toc,title,page]{appendix}
\usepackage{cite} 
\usepackage{amsfonts}
\usepackage{amsmath}
\usepackage{amssymb}
\usepackage{physics}
\usepackage[dvipsnames]{xcolor}
\usepackage{feynmp-auto}
\usepackage{epsfig}
\usepackage{slashed}
\allowdisplaybreaks
\usepackage{comment}
\usepackage{authblk}
\usepackage{blindtext}
\usepackage[colorlinks=true, allcolors=red]{hyperref}
\usepackage{setspace}
\usepackage[english]{babel}

\newcommand{\bc}{\begin{center}}
\newcommand{\ec}{\end{center}}
\newcommand{\me}{\medskip}
\newcommand{\ran}{\rangle}
\newcommand{\lan}{\langle}
\newcommand{\ra}{\rightarrow}

\newcommand{\nn}{\nonumber}

\def\rhol3{\rho^f_{\vb*l_3}}

\title{\textbf{Entanglement Entropy Distributions of a Muon Decay
}}
\author{Shanmuka Shivashankara, Patti Rizzo,  Nicole Cafe\footnote{sshivashankara@colgate.edu, prizzo@colgate.edu, ncafe@colgate.edu}}
\affil{\itshape\small Colgate University, Department of Physics and Astronomy, \\ \itshape\small Hamilton, NY  USA}
\date{
\small Competing interests: The authors declare there are no  competing interests.}

\numberwithin{equation}{section}

\begin{document}
\maketitle
%\clearpage

\begin{abstract}
Divergences that occur in density matrices of decay and scattering processes are shown to be regularized by tracing and unitarity or the optical theorem.  These divergences are regularized by the lifetime of the decaying particle or the total scattering cross section.  Also, this regularization is shown to give the expected helicities of final particles.  The density matrix is derived for the weak decay of a polarized muon at rest, $\mu^- \ra \nu_{\mu} (e^- \bar \nu_e)$, with Lorentz invariant density matrix entries and unitarity upheld at tree level.  The electron's von Neumann entanglement entropy distributions are calculated with respect to both the electron's emission angle and energy.  The angular entropy distribution favors an electron  emitted backwards with respect to the muon's polarization given a minimum volume regularization. The kinematic entropy distribution is maximal at half the muon's rest mass energy.  These results are similar to the electron's angular and kinematic decay rate distributions.  Both the density matrix and entanglement entropy can be cast either in terms of ratios of areas or volumes.
\end{abstract}

\section{Introduction}

Recently, the literature has applied the tools of quantum field theory to evaluating quantum information science (QIS) metrics for scalar and electromagnetic scattering \cite{Seki} - \cite{shiva}.  QIS metrics include the density matrix, von Neumann entanglement entropy, mutual information, etc.  The density matrix carries information about the particles in terms of degrees of freedom such as momenta and polarizations.  The entanglement entropy and mutual information are the degree of maximum knowledge and correlation between degrees of freedom, respectively.  \cite{Lykken} provides a review of QIS for particle physicists.\me

\cite{Seki} considers a scalar $\Phi^4$-interaction and a time dependent interaction.  \cite{Fan1} - \cite{Araujo2} consider an electromagnetic interaction, $e^-e^+\rightarrow \mu^-\mu^+$.  The Lorentz invariance of entanglement entropy is studied in \cite{Fan1}.  Entropies and mutual information are calculated in both the helicity and spin basis in \cite{Fan2}.  \cite{Fan3} evaluates two scattering processes being correlated through an entanglement of initial particles.  \cite{Araujo1,Araujo2} consider a witness particle or spectator.  During a scattering of two particles, one of these two is entangled with another particle (witness) that does not participate in the interaction.  The scattering alters the reduced density matrix of the witness.  This means the information of the witness particle is altered by the scattering event despite not participating directly in the interaction.  However, the latter is not true when including unitarity \cite{shiva}.  \cite{shiva} also evaluates QIS metrics for Compton scattering wherein the photon is entangled with a witness photon.  In spite of not interacting directly, the electron and witness photon acquire a non-zero mutual information, i.e. become entangled, after the Compton scattering.\me    

Whereas works \cite{Seki} - \cite{shiva} investigate scattering processes, QIS metrics are evaluated herein for a weak interaction $\mu^-\ra\nu_\mu(\bar{\nu_e}e^-)$.    When deriving the density matrix, this work is distinguished from  \cite{Seki} - \cite{Araujo2} by upholding unitarity up to tree level, i.e. using the optical theorem.  For this reason, the normalization of the density matrix is unchangeded by the interaction.  Also, unlike \cite{Seki} - \cite{shiva}, this work regularizes common divergences that occur in the density matrices of decay and scattering processes.  Although the density matrix is calculated for a decay process, the same techniques are readily applicable to scattering processes.  \me

In section \ref{densitymatrix}, the reduced density matrix of a massless electron for a muon decay is derived.  A common divergence appears in all terms and is regularized by tracing and the optical theorem.  Furthermore, the expected helicity of the electron provides confirmation of the regularization.  In section \ref{entdist}, Lorentz invariant density matrix elements are extracted from the electron's reduced density matrix of momenta.  These elements allow for calculating the von Neumann entanglement entropy distributions and total entropy.  The resulting angular and kinematic entropy distributions for the electron are peaked similarly with respect to the corresponding decay rate distributions.  An additive volume divergence occurs in the total entropy.  Section \ref{scat} gives the regularized density matrix for the annihilation process $e^-e^+\ra \sum_x x\bar x$.\me

\section{Density matrix of muon decay}\label{densitymatrix}

\begin{figure}[ht]
\bc
\includegraphics[width=\textwidth]{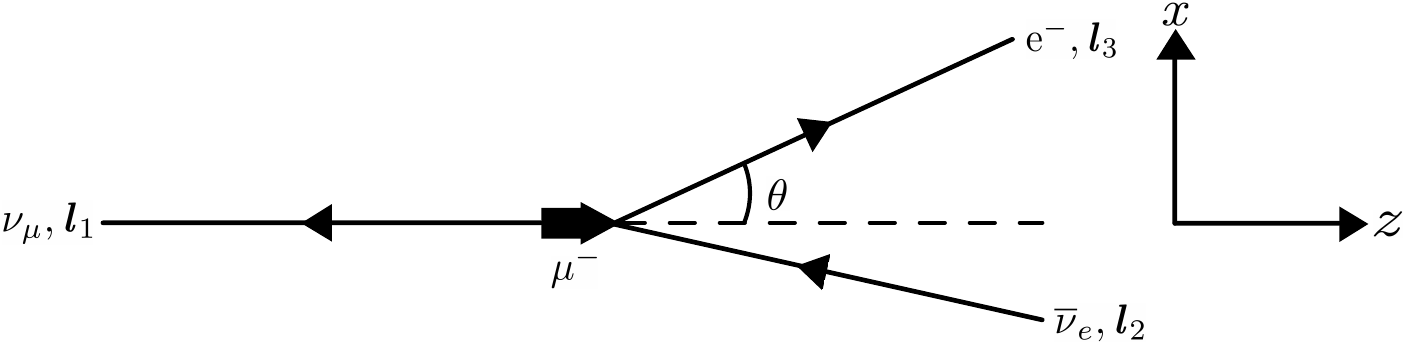}
\caption{A polarized muon at rest decays into three particles: $\mu^-\ra \nu_\mu(\bar{\nu}_e e^-)$.}
 \label{feynman}
\ec
\end{figure}

\noindent Suppose a muon is prepared in a pure state $|i\ran = |\vb*p, \uparrow \ran$ with momentum $\vb*p$ and spin up along the $+z$-axis (see Figure \ref{feynman}).  Its initial density matrix is  $\rho^i=|i\ran \lan i|$.  The $S$-matrix or unitarity operator  gives the final state $|f\ran = S |i\ran$ where $S=1+iT$ and $T$ is the transition operator.  The final state has a Fock space of 1 or 3 particles since the muon either does not decay or decays into $\nu_\mu(\bar{\nu}_e e^-)$.  A free 1-particle and 3-particle Hamiltonian have bases that span the final state, implying a direct sum of final Hilbert spaces $\mathcal{H}_\mu \oplus\mathcal{H}_{\nu_\mu} \otimes\mathcal{H}_{\bar{\nu}_e}\otimes\mathcal{H}_e$.   The three-particle state is written as
$|\vb*{l_1},s_1;\vb*{l_2},s_2;\vb*{l_3},s_3\rangle = |\vb*{l_1},s_1\rangle\otimes |\vb*{l_2},s_2\rangle\otimes|\vb*{l_3},s_3 \rangle,
$ where ($\vb*{l_i},s_i$) are the momentum-polarization pairs for the three particles.  The inner product of a state is
$\langle \vb*{p},r|\vb*{q},s\rangle$=
$2E_{\vb*{p}} (2\pi)^3\delta^{(3)}(\vb*p-\vb*q)\delta^{r,s}.$   
Defining $Q_{\vb*l s} \equiv \sum_{s}\ \int \frac{d^3\vb*{l}}{(2\pi)^32E_{\vb*{l}}}$, the single and three particle projection operators are written as
\begin{align*}
I_{1-particle} & = Q_{\vb*l s}\ |\vb*l,s\ran \lan \vb*{\vb*l},s|\ \ \text{and}\\
I_{3-particles} & =  \prod_{i=1}^3 Q_{\vb*{l_i}s_i} |\vb*{l_i},s_i\ran \lan \vb*{\vb*{l_i}},s_i|.
\end{align*}

These particle projection operators are placed next to $iT$ in the final state as follows.
\begin{align*}
|f\ran &= S|i\ran = (1 + i T) |i\ran\\
 &=|i\ran + (I_{1-particle} + I_{3-particles})i T |i\ran.
\end{align*}
This gives a final density matrix $\rho^f = | f \ran \lan f |$ or 
\begin{align}\label{rhof}
\rho^f  =& \indent | i \ran \lan i | \\
  & + \Big(Q_{kr}Q_{ls}\langle\vb*{l},s|i T|i\rangle\langle i|(-i T^\dagger)|\vb*{k},r\rangle \Big)\ | \vb*l,s\ran \langle\vb*{k},r|\nn \\
& + \prod^{3}_{m,n=1}Q_{{\vb*k_m}{r_m}}Q_{{\vb*l_n}{s_n}}\langle\vb*{k_m},r_m|i T|i\rangle\langle i|(-i T^\dagger)|\vb*{l_n},s_n\rangle|\vb*{k_m},r_m\rangle\langle \vb*{l_n},s_n|\nn \\
& + \Big(Q_{\vb*k r} \lan \vb*k,r | i T|i\ran\   |\vb*k,r \ran \lan i|+ h.c.\Big)\nn \\
& + \text{four other coherence terms.}\nn 
\end{align}

Apart from the matrix positions, e.g. $|i\ran \lan i|,\ |\vb*l,s\ran \lan \vb*k,r|$, etc., the  final density matrix, $\rho^f$, has Lorentz invariant entries.  After tracing over $\mu^-$, the reduced density matrix for $\nu_\mu(\bar{\nu}_e e^-)$ is  
\begin{align}\label{red3}
\rho^f_{\nu_{\mu}\bar{\nu_e}e}&=tr_{\mu}(\rho^f)\nn \\
    &=  2E_{\vb*p}V \Big(1-T \Gamma\Big)+\\
    &\indent \prod^{3}_{m,n=1}Q_{{\vb*k_m}{r_m}}Q_{{\vb*l_n}{s_n}}\langle\vb*{k_m},r_m|i T|i\rangle\langle i|(-i T^\dagger)|\vb*{l_n},s_n\rangle\ |\vb*{k_m},r_m\rangle\langle \vb*{l_n},s_n|.\nn
\end{align} 
  $\Gamma$ above is the muon's decay width.  It occurs from the fourth line in equation (\ref{rhof}) by tracing over  $\mu^-$ and using the optical theorem \cite{peskin}.  $V = (2\pi)^3 \delta^3(0)$ and  $T = 2\pi \delta(0)$ are the unregularized volume and time, respectively.  Whether calculating density matrices of scattering or decay processes, these unregularized factors are a common occurrence \cite{shiva}.  

  The plus sign in equation (\ref{red3}) should be interpreted as a direct sum of the one-particle and three-particle states.  The factor $(1-T\Gamma)$ in the first term can be interpreted as the probability for the muon $not$ to decay, which must be zero.  This implies that $T$ is the inverse of the muon's total decay width $\Gamma$.  Confirmation of $T$'s regularization is seen below when calculating the $e^-$'s helicity and the trace of the density matrix.  In short, setting the first term in equation (\ref{red3}) to zero gives the conditional density matrix for a polarized muon decay. Regardless of whether $(1-T\Gamma)$ is assumed to be zero,   $\rho^f_{\nu_\mu \bar{\nu}_e e}$'s normalization is $\lan i|i\ran = 2E_{\vb*p}V$ after tracing over the three remaining particles. Hence, the normalization is unaffected by the decay.  \me   

Tracing over the neutrinos, the normalized electronic reduced density matrix is
\begin{align}\label{rede}
\rho^f_{e} =&\  tr_{{\nu_\mu\bar{\nu_e}}} (\rho^f_{\nu_\mu\bar\nu_{e} e})\nn\\
=&\ \sum_{t_3}\frac{T}{2m_\mu}  (\prod^{3}_{m=1}Q_{\vb*l_m s_m})\ M_{\vb*p,\uparrow}^{\vb*{l_1},s_1;\vb*{l_2},s_2;\vb*{l_3},s_3} \Big(M_{\vb*p,\uparrow}^{\vb*{l_1},s_1;\vb*{l_2},s_2;\vb*{l_3},t_3}\Big)^\dagger*\\
    &\indent (2\pi)^4 \delta^{(4)}(p- \sum_{i=1}^3 l_i)\ \Big(\frac{|\vb*{l_3},s_3\rangle\langle \vb*{l_3},t_3|}{2E  V}\Big),\nn
\end{align}
where $E=l_3$ is the massless $e^-$'s energy, $M_{\vb*p,\uparrow}^{\vb*{l_1},s_1;\vb*{l_2},s_2;\vb*{l_3},s_3}$ is the Feynman amplitude for the $\mu^-$ decay, and the transition amplitude is related to the Feynman amplitude by
\begin{align}
\langle \vb*{l_1},s_1; \vb*{l_2},s_2; \vb*{l_3},s_3|i T|i\rangle& = iM_{\vb*p,\uparrow}^{\vb*{l_1},s_1;\vb*{l_2},s_2;\vb*{l_3},s_3}\ (2\pi)^4 \delta^{(4)}(p- \sum_{i=1}^3 l_i).  \nn
\end{align}
The above electronic reduced density matrix lacks purity since $\rho^f_e \neq (\rho^f_e)^2$.  Its trace is 
\begin{align*}
tr(\rho^f_e) = T\Gamma = 1.
\end{align*}
Just as $\rho^f$ above, $\rho^f_e$'s matrix elements, i.e. the cofficients of $\dfrac{|\vb*{l_3},s_3\rangle\langle \vb*{l_3},t_3|}{2E V}$, are Lorentz invariant.  This follows from $T/2m_\mu = \dfrac{VT}{2m_\mu V}$, $Q_{\vb*l_m,s_m}$, the Feynman amplitude, and $(2\pi)^4\delta^4(\cdot)$ all being Lorentz invariant.\me

Assume in equation (\ref{rede}), the electron's polarization is its helicity, $\lambda$.  In the massless limit for the electron in weak interactions, its helicity is $\lambda=-1$ and the $\rho^f_e$'s polarization-coherence terms are zero.  The helicity-reduced density matrix follows by tracing equation (\ref{rede}) over $\vb*l_3$, giving
\begin{align}\label{redh}
\rho^f_{\lambda}= tr_{\vb*l_3}(\rho^f_e) = T\Gamma\ |\lambda=-1 \ran \lan \lambda=-1 |.
\end{align}
The latter correctly gives the electron's helicity, 
\begin{align*}
\lan \sigma_z \ran = tr(\sigma_z \rho^f_{\lambda}) = -T\Gamma = -1,
\end{align*} where $\sigma_z$ is Pauli's matrix.  
Conversely, the electron's helicity requires $T=\dfrac{1}{\Gamma}$.  As another example, consider the decay $\pi^- \ra e^- \bar \nu_e, \mu^- \bar \nu_\mu$.  Using the above techniques, the expected antineutrino helicity is 
\begin{align*}
T \sum_{x=e^-,\mu^-} \lambda_{\bar \nu_x}  \Gamma_{\pi^- \ra x \bar \nu_x} = 1   
\end{align*}
\me 
since the antineutrino's helicities are $+1$.  Therefore, again, $T$ must be the inverse of the total decay width.

Tracing equation (\ref{rede}) over the electronic polarizations gives a diagonal reduced density matrix of the electon's momenta.
\begin{align}\label{redm}
\rhol3 &= tr(\rho_e^f)\nn \\
&=T \int\dfrac{d^3\vb*l_3}{(2\pi)^3 2E}\ f(E,\theta) \dfrac{|\vb*l_3 \ran \lan \vb*l_3|}{2E V},\text{ where}\\
f(E ,\theta) &= \dfrac{G_F^2 m}{3\pi}\ E\ (3m-4E + \cos\theta\ (m-4E)),\nn
\end{align}
$G_F = 1.17\text{x}10^{-5}GeV^{-2}$ is the Fermi coupling constant, $E=l_3$ is the massless electron's energy, and $\theta$ is the electron's scattering angle with respect to the muon's polarization \cite{peskin}.

The trace of any density matrix being one requires upholding unitarity.  This follows from $tr(\rho) = tr (|f\ran \lan f|) = \lan f|f \ran = \lan i|S^\dagger S|i \ran=1$ for a normalized initial state $|i\ran$ and a unitary $S$-matrix.  The trace of equation (\ref{redm}) is $1.0044$ due to the muon's lifetime, $T=2.1969811\text{x}10^{-6}s$ \cite{pdg}, and the calculation being at tree level.  Hence, equation \ref{redm} approximately upholds unitarity  since radiative corrections and other final particle states, e.g. photon emission or pair production, were not included.\me

\section{von Neumann entanglement entropy distributions }\label{entdist}

Having the reduced density matrices for the electron in equations (\ref{rede}-\ref{redm}), other quantum information metrics can be calculated.  Since $\rho^f_\lambda$ is pure, only the momentum contributes to the massless electron's von Neumann entanglement entropy,
\begin{align*}
S^{EE}_{e}&=-tr\rho^f_e \log \rho^f_e = -tr\rho^f_\lambda \log \rho^f_\lambda - tr\rhol3 \log\rhol3\\
&= -tr\rhol3 \log \rhol3\\
&\equiv S^{EE}_{\vb*l_3}.
\end{align*}
The mutual information of the electron's momentum and helicity is
\begin{align*}
I &= S^{EE}_{\vb*l_3} + S^{EE}_{\lambda} - S^{EE}_e\\
&= 0.
\end{align*}
This shows momentum and helicity are not correlated as expected for weak interactions.  

The trace operator in the von Neumann entropy above, $S^{EE}_e$, follows from a continuous limit of the counting of states and recalling that $(2\pi\hbar)^3$ is the volume of phase space per state, 
\begin{align}\label{sumcont}
\sum_n \ra \dfrac{V}{(2\pi\hbar)^3}\int d^3 \vb*l_3 = \delta^3(0)\int d^3 \vb*l_3. 
\end{align}
In the last equality, the quantity $V=(2\pi\hbar)^3 \delta^3(0)$ in non-natural units was used.  Equation (\ref{sumcont}) is also implied by tracing the identity operator for 1-particle states, giving the total number of states.  By extracting $\delta^3(0)\int d^3\vb*l_3$ from equation (\ref{redm}) and including $\hbar$ and $c$, the matrix elements of $\rhol3$ are 
\begin{align}\label{matrix}
\dfrac{T f(E ,\theta)\hbar^2 c^3} {2E V}
\end{align} 
or a ratio of the ostensible partial volume, $\varrho (E,\theta) \equiv \dfrac{T f(E ,\theta)\hbar^2 c^3} {2E}$,
to the total accessible volume, $V$.  The cube root of the maximum partial volume is on the order of the muon's Compton wavelength, i.e.  $\varrho^{1/3}_{max} \sim \dfrac{h}{m_\mu c} \sim 10^{-14}m$.
With the partial volume, the electron's entanglement entropy distributions with respect to momentum and emission angle are calculated below and found to be similar to decay rate distributions.  Assume log base $e$ for the caculations below.  Alternatively, $\varrho/V$ can be cast as a ratio of areas, i.e. $a(E,\theta)/A$ with $A\equiv \dfrac{V}{cT}$ and $a(E,\theta) \equiv f(E,\theta)(\hbar c)^2/(2E)$.  Using equation (\ref{sumcont}), the density matrix in equation (\ref{redm}) can be rewritten in terms of ratios of areas or volumes.

\begin{align*}
\rhol3 &=\Big(\dfrac{V}{(2\pi\hbar)^3}\int d^3 \vb*l_3\Big)\  \dfrac{a(E,\theta)}{A}\  \dfrac{|\vb*l_3 \ran \lan \vb*l_3|}{2E V}\nn\\
&=\Big(\dfrac{V}{(2\pi\hbar)^3}\int d^3 \vb*l_3\Big)\  \dfrac{\varrho(E,\theta)}{V}\  \dfrac{|\vb*l_3 \ran \lan \vb*l_3|}{2E V}.\nn\\
\end{align*}

Using equation (\ref{matrix}), the daughter electron's von Neumann entanglement entropy for the polarized muon decay is  
\begin{align}\label{se-natural}
S^{EE}_e&=-tr(\rhol3 \log\rhol3)\nn\\
&=-\frac{V}{(2\pi\hbar)^3}\int d^3\vb*l_3\ \Big(\dfrac{\varrho(E,\theta)}{V} \log (\dfrac{\varrho(E,\theta)}{V})\Big).
\end{align}
The above equation looks similar to the Shannon entropy $\sum_i p_i \log p_i$.  
\begin{comment}
This suggests $\varrho(E,\theta)$ has a variance $\sigma_\varrho^2 = E[(\varrho(E,\theta) - \bar \varrho(E,\theta))^2]$. Let the regularization of $V$ lie in a range $1.34 \varrho(E,\theta)_{max} \leq V \leq (\bar \varrho(E,\theta) + 100\sigma_{\varrho})$ or $1.1\text{x}10^{-41}\leq V \leq 9.7\text{x}10^{-41}$ in cubic meters.      
\end{comment}
After simplifying equation (\ref{se-natural}), the electron's entanglement entropy becomes 
\begin{align}\label{se}
S^{EE}_e =-\dfrac{1}{(2\pi)^3 \Gamma} \int d\Omega \int dE\ \dfrac{E f(E,\theta)} {2} \log (\dfrac{ f(E,\theta) (\hbar c)^3} {2\Gamma E V}),
\end{align}
where $\Gamma = \hbar/T$ is the muon's decay width. $\Gamma$ and $S^{EE}_e$ are related by
\begin{align*}
S^{EE}_e = -\int d\Omega \int dE\ \dfrac{1}{\Gamma}\dfrac{d^2\Gamma}{d\Omega dE} \log \varrho(E,\theta) + \log V.
\end{align*}
In the above equation, the divergence in entropy, or $\log V$, is additive.  
For an unregularized $V$,  changes in entropy between different decay processes will yield finite results.
A larger regularized volume corresponds to a higher total entropy, $S^{EE}_e$. 

 \begin{figure}[htpb]
\begin{subfigure}{\textwidth}
\bc
\includegraphics[width=.8\textwidth]{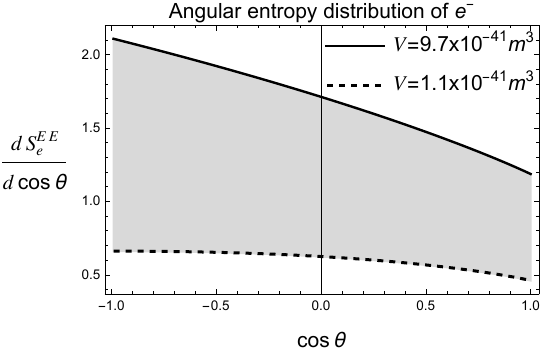}
 \caption{$\theta$ is the emission angle of the electron with respect to the muon's polarization. $V$ is the total volume accessible to the electron.  For the distribution to be peaked at $x=-1$, the minimum regularized volume is $V = 1.1\text{x}10^{-41}m^3$.  As $V$ grows, the distribution becomes proportional to its decay rate counterpart $\dfrac{d\Gamma}{d\cos \theta}$.  The plot's sample regularization range is $1.1\text{x}10^{-41}m^3 \leq V \leq 9.7\text{x}10^{-41}m^3$. \vspace{.2cm} }
  \label{dscos}
  \ec
\end{subfigure}
\begin{subfigure}{\textwidth}
\bc
\includegraphics[width=.80\textwidth]{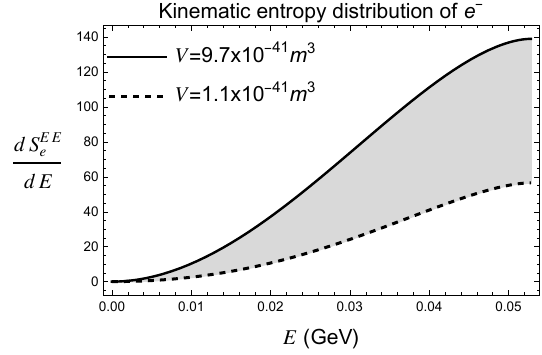}
  \caption{The massless $e^-$'s energy is $0\leq E \leq \dfrac{m_\mu}{2}$, where $m_\mu$ is the $\mu^-$'s rest mass.}
  \label{dse}
  \ec
\end{subfigure}
\caption{}
\label{entropy-dists}
\end{figure}
Differentiating equation (\ref{se}) gives the electron's angular entropy distribution $\dfrac{d S^{EE}_e}{d\cos\theta}$ and kinematic entropy distribution $\dfrac{d S^{EE}_e}{dE}$, which are long equations and not written herein. See Figures \ref{dscos} and \ref{dse}.  For regularizations $V\geq 1.34\varrho_{max}=1.1\text{x}10^{-41}m^3$, the angular entropy distribution peaks at an electronic emission antiparallel with respect to the muonic polarization as does the decay rate distribution $\dfrac{d\Gamma}{d\cos\theta}$.  

The kinematic entropy distribution peaks at an energy of $E=m_\mu/2$ as does the decay rate distribution $\dfrac{d\Gamma}{dE}$.  For a sufficiently large $V$, the entropy distributions are proportional to their decay rate counterparts.  The total entropy is $  S^{EE}_e \geq .91$ for $V \geq 1.34 \varrho_{max}$.

\section{Regularization of a scattering process}\label{scat}

For decay processes, evaluation of density matrices require relating $T =2\pi \delta(0)$ with the total decay width via $T=1/\Gamma$. In other words, $T$ is the muon's lifetime.  For the density matrix of a scattering process, $V/T$ should be related to the total accessible scattering cross section as shown below.  

Consider the high energy annihilation process $e^-e^+\ra \sum_x x\bar x$ in the CoM frame where $e^-,e^+$ have particular helicities and $x,\bar x$ is a particle, antiparticle pair.  Following the procedure of section \ref{densitymatrix} and using the optical theorem, the final reduced density matrix of final particle $x$ equals the direct sum

\begin{equation}\label{scatt}
\begin{split}
\rho^f_{e^-,\mu^-,\tau^-,etc.} =& \Big(1-\dfrac{\sum_f \sigma(e^-,e^+ \ra f)}{\dfrac{V}{T |\boldsymbol{\upsilon_{e^-,e^+}}|}}\Big) \ \oplus\\ 
& \dfrac{1}{\dfrac{V}{T |\boldsymbol{\upsilon_{e^-,e^+}}|}}\ \  \sum_{\lambda_x,\lambda'_x}\ \sum_{x=e^-, \mu^-, \tau^-, etc.} \int d^3\vb*l_x\dfrac{d^3\sigma^{\lambda_x,\lambda_x^{'}}_{e^-,e^+\ra x\Bar x}}{d^3\vb*l_x}\ \dfrac{|\vb*l_x, \lambda_x \ran \lan \vb*l_x,\lambda_x^{'} |}{2E_{\vb*l_x}V}. 
\end{split}
\end{equation}
$\vb*l_x$ and $\lambda_x$ above are final particle $x$'s momentum and helicity, respectively.  $\vb*{\upsilon_{e^-,e^+}}$ is the relative velocity of the initial particles.  
The differential cross section matrix elements in the second term are
\begin{align*}
\dfrac{d^3\sigma^{\lambda_x,\lambda_x^{'}}_{e^-,e^+\ra x\Bar x}}{d^3\vb*l_x} &\equiv \frac{1}{2E_{e^-} 2E_{e^+}|\vb*v_{e^-,e^+}|}* \\ (\dfrac{1}{(2\pi)^3 E_{\vb*l_x}}Q_{\vb*l_{\bar x} \lambda_{\bar x}})\ M_{e^-,e^+}^{\vb*{l_x},\lambda_x;\vb*{l_{\bar x}},\lambda_{\bar x}} &\Big(M_{e^-,e^+}^{\vb*{l_x},\lambda'_x;\vb*{l_{\bar x}},\lambda_{\bar x}}\Big)^\dagger
     (2\pi)^4 \delta^{(4)}(p-  l_x - l_{\bar x}).
\end{align*}
Note that the initial particles and final particles occupy different 2-particle vector spaces.  Thus, the initial and final electrons are traced separately.  The first term in equation (\ref{scatt}) is from tracing over the initial particles $e^-, e^+$.  For unitarity to hold exactly or $tr(\rho)=1$, all additional final particle states should be added to equation \ref{scatt}, e.g. including Bremsstrahlung.\\  

The density matrix has Lorentz invariant entries.  e.g., the ratio 
$\dfrac{\sum_f \sigma(a,b\ra f)}{\dfrac{V}{T|\boldsymbol {\upsilon_{a,b}}|}}$ in the first term in equation (\ref{scatt}) is manifestly Lorentz invariant \cite{peskin} when rewritten as 
\begin{center}$\dfrac{2E_a 2E_b |\boldsymbol {\upsilon_{a,b}}|\sum_f \sigma(a,b\ra f)}{\Big(\dfrac{2E_aV\ 2E_bV}{VT}\Big)}$. 
\end{center}
The first term in equation (\ref{scatt}) is the probability for no scattering to occur.  Setting this term to zero gives the conditional density matrix for $e^-,e^+$ scattering.  This forces the regularization for scattering to be
\begin{equation}\label{reg1}
\dfrac{V}{T|\boldsymbol{\upsilon_{e^-,e^+}}|} = \sum_f \sigma(e^-,e^+\ra f)\equiv \sigma_T.
\end{equation}

With the latter regularization, the expected helicity of $x$ from equations (\ref{scatt}) and (\ref{reg1}) is 
\begin{equation}
\begin{split}
    \lan \sigma_z \ran & = tr(\sigma_z\otimes \rho_x)\\
    & = \sum_x \sum_{\lambda_x} \lambda_x \dfrac{\sigma^{\lambda_x,\lambda_x}_{e^-,e^+ \ra x \Bar{x}}}{\sigma_{T}}.
    \end{split}
\end{equation}
This is a weighted average of the helicities with each weight being the probability of a particular scattering process.\\

\section{Discussion}

Herein, the density matrix and  von Neumann entanglement entropy, $S^{EE}$, were derived for the decay of a polarized muon at rest, $\mu^- \ra \nu_\mu (\bar{\nu_e} e^-)$.  Unitarity is upheld at tree level.  The divergences in the density matrix for a decay and scattering process are resolved in sections \ref{densitymatrix} and \ref{scat} by the optical theorem and tracing.  Alternatively, knowing the expected helicities of final particles suggests the regularization.\me  

When evalulating $S^{EE}$, an unregularized volume, $V$, appears.  This volume is additive in the total entropy. If this accessible volume is not regularized and the same across all interactions, the differences in entropies across a variety of decays will be finite.  Otherwise, an appropriate constraint for regularization is needed.  For a regularized volume exceeding the cube of the muon's Compton wavelength, angular and kinematic entropy distributions for the electron are found to be similar to the decay rate distributions.  The common appearance of ratios of areas or volumes in the density matrix and entanglement entropy for both decay and scattering processes requires further investigation.\me

Future work should compare the entanglement entropies or distributions for different decay processes involving scalar, electromagnetic, weak, or strong interactions.  e.g., the negative pion has mainly two decay modes, $\pi^- \ra \mu^- \bar{\nu_\mu}$ or $e^- \bar{\nu_e}$.  Helicity suppression can be reinterpreted as entropy suppression for the electronic mode. \me

\section{Acknowledgements}

SS thanks Ms. Abigail Wiinikainen for initial work on this project.  This work was partly funded by a Colgate University Research Council grant.

\end{document}